\documentclass[
 reprint,
 amsmath,amssymb,
 aps,
 prb,
 floatfix
]{revtex4-2}

\usepackage[dvipdfmx]{graphicx}% Include figure files
\usepackage{dcolumn}% Align table columns on decimal point
\usepackage{bm}% bold math
\usepackage{braket}
\usepackage{here}
\usepackage{ulem}
\usepackage[usenames]{xcolor}

\begin{document}

\preprint{APS/123-QED}

\title{Optical Hall response in spin-orbit coupled metals:\\ Comparative study of magnetic cluster monopole, quadrupole, and toroidal orders}% Force line breaks with \\

\author{Tatsuki Sato$^1$}
\author{Yuma Umimoto$^1$}
\author{Yusuke Sugita$^2$}
\author{Yasuyuki Kato$^2$}
\author{Yukitoshi Motome$^2$}
\affiliation{%
	$^1$Department of Advanced Materials Science, University of Tokyo, Kashiwa 277-8561, Japan \\
	$^2$Department of Applied Physics, University of Tokyo, Tokyo 113-8656, Japan
}%

\begin{abstract}
The optical Hall response is theoretically studied for spin-orbit coupled metals with ferroic orders of cluster-type magnetic multipoles.
We find that different magnetic multipoles give rise to distinct spectra in the optical Hall conductivity.
In the cases of monopole and quadrupole orders, the optical Hall response appears predominantly in high- and low-energy regions, which correspond to the energy scales of electron correlation and kinetic energy, respectively, while
the response is dispersed and rather weak in the case of toroidal order.
By decomposing the spectra into different interband contributions, we reveal selection rules stemming from the interplay between the antisymmetric spin-orbit coupling and the underlying multipoles.
Our results suggest that the optical Hall measurement is useful to detect and distinguish the cluster-type magnetic multipole orders.
\end{abstract}

\maketitle

\section{Introduction}
\label{sec:Introduction}
In condensed matter physics, multipoles provide a key concept to understand physical properties through classification of spatial distributions of electrons with charge, spin, and orbital degrees of freedom.
Typical examples are the multipoles defined in the atomic scale, which have been used to characterize electronic and magnetic phases in $f$-electron systems~\cite{Kuramoto2009, Santini2009}.
Multipoles can also be defined in an extended scale over several atomic sites and such cluster-type extensions have recently garnered great attention as a source of intriguing phenomena~\cite{Spaldin2007, Spaldin2008, Yanase2014, Hayami2014, Suzuki2017, Gao2018, Hayami2018, Watanabe2018, Suzuki2019}.
For instance,
a magnetic toroidal dipole induces the second harmonic generation in LiCoPO$_{4}$~\cite{Aken2007}, and the magnetocurrent effect in UNi$_4$B~\cite{Hayami2014, Saito2018} and Ce$_3$TiBi$_5$~\cite{Shinozaki2020},
a magnetic octupole plays a crucial role in the anomalous Hall effect~\cite{Nakatsuji2015, Suzuki2017},
the anomalous Nernst effect~\cite{Ikhlas2017},
and the magneto-optical Kerr effect~\cite{Higo2018} in Mn$_{3}$Sn,
and a magnetic quadrupole causes the magnetoelectric effect in $A$(TiO)Cu$_4$(PO$_4$)$_4$ ($A$=Ba, Sr, and Pb)~\cite{Kimura2016, Kato2017, Kimura2018, Kato2019}.
Thus, it is useful to identify relevant multipoles for predicting the electronic, magnetic, transport, and optical properties.
At the same time, measurement of these properties enables us to identify the relevant multipoles.
For instance, in the linear response, the diagonal, traceless symmetric, and antisymmetric parts of the magnetoelectric effect have one-to-one correspondence with the magnetic monopole, quadrupole, and toroidal dipole, respectively~\cite{Spaldin2008}.
Such studies have been extensively performed in the DC limit for metallic systems.
Although the AC responses would also serve as useful tools as shown for the optical responses in insulating materials, they have not been studied systematically thus far.

In the present study, we theoretically study the optical Hall response for different types of cluster-type magnetic multipoles.
For a minimal model defined on a layered structure where spatial inversion symmetry is broken at each lattice site, we compute the electronic band structure and the optical Hall conductivity in the presence of ferroic orders of the cluster-type magnetic monopole, quadrupole, and toroidal dipole (toroidal).
We reveal that despite the similarity in the band structure, the optical Hall responses exhibit distinct frequency dependence for the three types of the multipoles.
By decomposing the responses into the interband contributions and analyzing them with the atomic bases, we show that the distinct behaviors can be understood from optical selection rules arising from the interplay between the antisymmetric spin-orbit coupling and the coupling of electrons to the underlying multipole orders.

This paper is organized as follows.
In Sec.~\ref{sec:Model}, we introduce the model with the cluster-type magnetic multipole orders.
We present the results for the electronic band structure in Sec.~\ref{sec:ElectronicBandStructure} and the optical Hall conductivity in Sec.~\ref{sec:OpticalHallConductivity}.
From the decomposition into the interband contributions, we find optical selection rules in Sec.~\ref{sec:InterbandContribution}.
In Sec.~\ref{sec:Discussion}, we discuss the origin of the optical selection rules by applying the perturbation theory in the atomic limit.
Section~\ref{sec:Summary} is devoted to the summary.
We also study a variant of the model in Appendix to confirm the generality of the optical selection rules.

\section{Model}
\label{sec:Model}

\begin{figure}[t]
	\centering
	\includegraphics[width = 0.8\columnwidth]{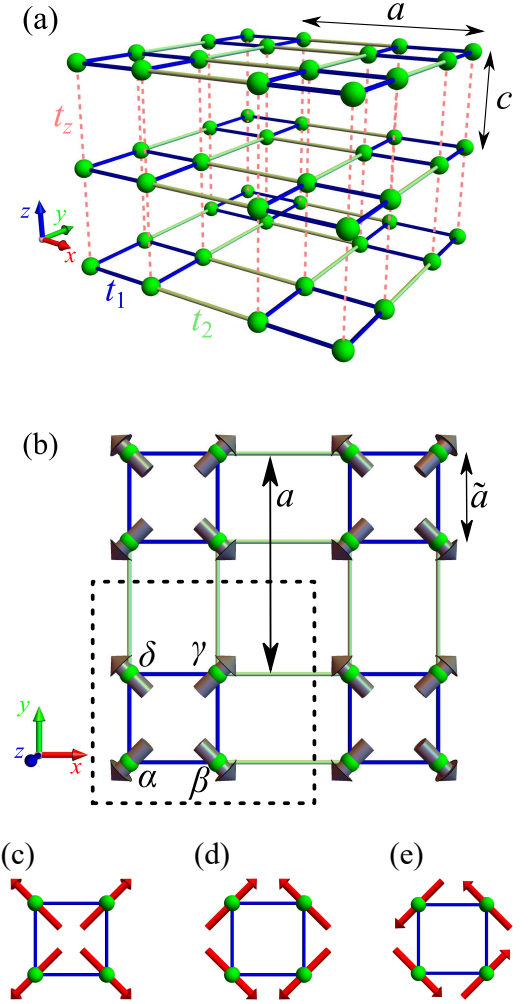}
	\caption{
		\label{fig:Structure_Square}
		Schematics of (a) a perspective view and (b) a top view of the layered square-cluster lattice.
		In (a), the transfer integrals $t_1$, $t_2$, and $t_z$ are shown.
		$a$ and $c$ are the lattice constants in the $xy$ plane and in the $z$ direction, respectively.
		In (b), $\tilde{a}$ represents the size of the square cluster.
		The dotted square indicates a unit cell (a square cluster) with four sublattices $\alpha$, $\beta$, $\gamma$, and $\delta$.
		The gray arrows denote the directions of ${\bm D}_{l}$ at each site; see Eqs.~(\ref{eq:D_l}) and (\ref{eq:theta^D_l}).
		(c)-(e) Schematics of cluster-type magnetic multipoles composed of four magnetic dipoles (red arrows): (c) monopole (d) quadrupole, and (e) toroidal.
	}
\end{figure}

We consider a minimal model with ferroic orders of cluster-type magnetic multipoles.
We adopt a single-band tight-binding model on a layered lattice structure, where each layer consists of a periodic array of four-site square clusters, as shown in Fig.~\ref{fig:Structure_Square}(a).
Each square cluster can accommodate magnetic cluster multipoles composed of four spins, such as monopole, quadrupole, and toroidal, as shown in Figs.~\ref{fig:Structure_Square}(c), \ref{fig:Structure_Square}(d), and \ref{fig:Structure_Square}(e), respectively.
Note that spatial inversion symmetry is broken at each lattice site, while it is retained at the centers of square plaquettes in each layer and of cuboids defined by two squares in adjacent layers.
A similar model with hexagonal clusters was discussed in the previous study~\cite{Hayami2014} (see
Appendix).
The Hamiltonian of our model is given by
\begin{align}
	{\mathcal H}
	=
  {\mathcal H}_{t}
  + {\mathcal H}_\textrm{ASOC}
  + {\mathcal H}_\textrm{MF}
  + {\mathcal H}_\textrm{Zeeman} ,
	\label{eq:Hamiltonian_All}
\end{align}
where
\begin{align}
  \mathcal H_{t}
  &=
  - \sum_{i, j} \sum_{\sigma} t_{i j} (c^\dagger_{i \sigma}c_{j \sigma} + \textrm{h}.\textrm{c}.) ,
  \label{eq:H_t}\\
  \mathcal H_\textrm{ASOC}
	&=
	2 \sum_{\bm k, l} [ {\bm s}_{{\bm k} l} \times {\bm D}_{{\bm k} l} ]_{z} ,
	\label{eq:H_ASOC}\\
  \mathcal H_\textrm{MF}
  &=
  - \sum_{i} \bm M _i \cdot {\bm s}_{i} ,
  \label{eq:H_MF}\\
	\mathcal H _\textrm{Zeeman} %&
	&= - {\bm B} \cdot \sum_{i} {\bm s}_{i} .
	\label{eq:H_Zeeman}
\end{align}

$\mathcal H_t$ in Eq.~(\ref{eq:H_t}) describes the hoppings of electrons.
$c^{\dagger}_{i \sigma}$ ($c_{i \sigma}$) is the creation (annihilation) operator for an electron at site $i$ with spin $\sigma$.
We take into account three types of transfer integrals between neighboring sites:
the intralayer ones $t_1$ and $t_2$ within and between the clusters, respectively,
and the interlayer one $t_z$ [see Fig.~\ref{fig:Structure_Square}(a)].
All of the other transfer integrals between further-neighbor sites are assumed to be zero.

$\mathcal H_{\rm ASOC}$ in Eq.~(\ref{eq:H_ASOC}) describes the antisymmetric spin-orbit coupling.
It originates from the interplay among the atomic spin-orbit coupling, off-site orbital hybridization, and the crystalline electric field~\cite{Hayami2014}.
${\bm s}_{{\bm k} l}$ is the Fourier transform of the spin operator at site $i$, ${\bm s}_{i}=\frac{1}{2} \sum_{\sigma, \sigma'} c_{i \sigma} ^\dagger \bm \sigma_{\sigma \sigma'} c _{i \sigma'}$, where ${\bm \sigma}$ is the vector of the Pauli matrices; ${\bm k}$ and $l$ denote momentum and sublattice, respectively.
${\bm D}_{{\bm k} l}$ represents a sublattice-dependent vector antisymmetric with respect to $k_{z}$, which is given by
\begin{align}
	{\bm D}_{{\bm k} l} = {\bm D}_l \sin(k_{z} c),
	\label{eq:D_kl}
\end{align}
with
\begin{align}
	{\bm D}_{l}
	=
	D \ ( \cos \theta^D_l, \sin \theta^D_l, 0) .
	\label{eq:D_l}
\end{align}
Here, $D$ is the coupling constant and
\begin{align}
	\theta^D_l
	=
	\frac{\pi}{2} n_l - \frac{3\pi}{4},
	\label{eq:theta^D_l}
\end{align}
where $n_l = $ 0, 1, 2, and 3 correspond to the sublattices $l = \alpha$, $\beta$, $\gamma$, and $\delta$, respectively.
The directions of ${\bm D}_l$ are shown by the gray arrows in Fig.~\ref{fig:Structure_Square}(b).
The $k_z$ dependence in Eq.~(\ref{eq:D_kl}) comes from the off-site orbital hybridization along the $z$ axis,
where $c$ is the lattice constant in the $z$ direction~\cite{Hayami2014}.

$\mathcal H_{\rm MF}$ in Eq.~(\ref{eq:H_MF}) describes the exchange coupling between itinerant electron spins and magnetic multipoles at the mean-field level.
The multipoles are composed of magnetic moments ${\bm M}_i$, which can be regarded as localized moments coupled to itinerant electrons or mean fields decoupled from the Coulomb interaction between itinerant electrons.
We assume ferroic orders of three types of cluster-type magnetic multipoles, monopole, quadrupole, and toroidal [see Figs.~\ref{fig:Structure_Square}(c)-\ref{fig:Structure_Square}(e)].
Then, ${\bm M}_i$ depends only on the sublattice $l$ as
\begin{align}
	{\bm M}_l = M(\cos \theta^M_l, \sin \theta^M_l, 0),
	\label{eq:M_l}
\end{align}
where $M$ denotes the magnitude of the magnetic moments and
\begin{align}
	  & \theta^M_l =   \frac{\pi}{2} n_l - \frac{3 \pi}{4} \quad{\rm for \ monopole ,}
	\label{eq:theta^M_l_flux} \\
	  & \theta^M_l = - \frac{\pi}{2} n_l - \frac{\pi}{4} \quad{\rm for \ quadrupole , }
	\label{eq:theta^M_l_quadrupole} \\
	  & \theta^M_l =   \frac{\pi}{2} n_l - \frac{\pi}{4} \quad{\rm for \ toroidal.}
	\label{eq:theta^M_l_toroidal}
\end{align}

$\mathcal H_{\rm Zeeman}$ in Eq.~(\ref{eq:H_Zeeman}) represents the Zeeman coupling to an external magnetic field ${\bm B}$.
We assume that the magnetic field couples only to the electron spins and neglect a canting of the magnetic moments ${\bm M}_{i}$ by ${\bm B}$.

\section{Result}
\label{sec:Result}

In this section, we present the results of the electronic and transport properties for the model in Eq.~(\ref{eq:Hamiltonian_All}) in the presence of ferroic orders of the cluster-type magnetic multipoles.
In Sec.~\ref{sec:ElectronicBandStructure}, we show that the electronic band structures look similar to each other for the monopole, quadrupole, and toroidal orders.
Despite the similarity, however, we show that the frequency dependences of the optical Hall conductivity are substantially different in Sec.~\ref{sec:OpticalHallConductivity}.
In Sec.~\ref{sec:InterbandContribution}, to discuss the origin of the differences, we analyze the contributions from different interband processes.
All the following calculations in this section are obtained for the model parameters, $a = c = 1$, $\tilde a = 0.35$, $t_1 = 1.25$,  $t_2 = 0.75$, $t_z = 1$, $D = 0.5$, $M = 8$, and ${\bm B}=(0.5,0,0)$, which correspond to the strongly correlated metal under a small magnetic field.

\subsection{Electronic band structure}
\label{sec:ElectronicBandStructure}

\begin{figure}[t]
	\centering
	\includegraphics[width = 0.9 \columnwidth]{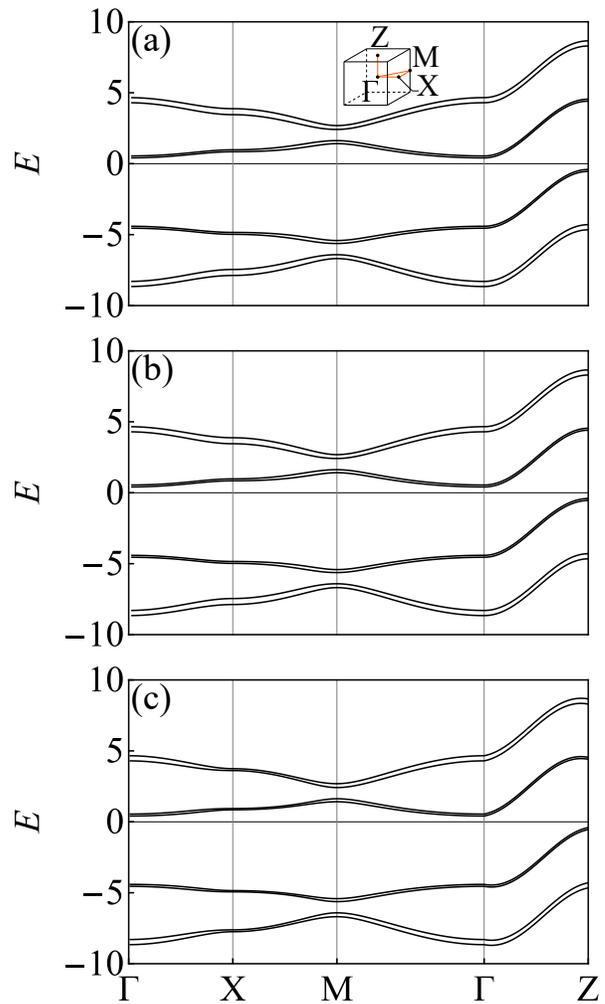}
	\caption{
		\label{fig:Band_Square}
		Electronic band structures of the model in Eq.~(\ref{eq:Hamiltonian_All})
		in the presence of the ferroic order of (a) monopole, (b) quadrupole, and (c) toroidal.
		The inset of (a) shows the first Brillouin zone and the red lines indicate the symmetric lines on which the electronic band structures are presented.
    The results are obtained with $a = c = 1$, $\tilde a = 0.35$, $t_1 = 1.25$, $t_2 = 0.75$, $t_{z} = 1$, $D = 0.5$, $M = 8$, and ${\bm B}=(0.5, 0, 0)$.
	}
\end{figure}

Figure~\ref{fig:Band_Square} shows the electronic band structures of the model in Eq.~(\ref{eq:Hamiltonian_All}) in the presence of ferroic orders of the cluster-type magnetic multipoles: (a) monopole, (b) quadrupole, and (c) toroidal.
In all cases, we obtain eight bands corresponding to the four sublattices and spin degrees of freedom.
The eight bands are split into two groups by the exchange coupling to the magnetic multipoles, $\mathcal H_\textrm{MF}$ in Eq.~(\ref{eq:H_MF}):
four lower(higher)-energy bands correspond to the bands with spins $\bm s_i$ (anti)parallel to the magnetic moments $\bm M_i$.
Further splitting of each four into two groups is brought by $\mathcal H_{t}$ in Eq.~(\ref{eq:H_t}) and the smallest splitting is caused by $\mathcal H _\textrm{Zeeman}$ in Eq.~(\ref{eq:H_Zeeman}) (see the discussion in Sec.~\ref{sec:Discussion}).
The overall band structures are similar to each other for the three types of multipoles, but nevertheless they lead to distinct optical Hall responses as shown in the next subsection.
We note that the band bottom is shifted along the $k_{z}$ direction ($\Gamma$-Z) in the case of the toroidal order, which is parallel to the toroidal moment, as shown in Fig.~{\ref{fig:Band_Square}}(c)~\cite{Yanase2014, Hayami2014},
whereas no such a shift is seen for the monopole and quadrupole orders as shown in Figs.~{\ref{fig:Band_Square}}(a) and \ref{fig:Band_Square}(b), respectively.

\subsection{Optical Hall conductivity}
\label{sec:OpticalHallConductivity}

We calculate the optical Hall conductivity $\sigma^{\mu z} (\omega)$ for an electric current in the $\mu$ direction induced by that in the $z$ direction.
It is obtained by using the Kubo formula as
\begin{align}
	\sigma^{\mu z} (\omega)
	=
	\sum_{m, n} \sigma^{\mu z}_{m,n}(\omega) ,
	\label{eq:KuboWhole}
\end{align}
where
\begin{align}
	\sigma^{\mu z}_{m,n}(\omega)
	= &
	\sum_{\bm k}
	\frac{e^2}{\hbar}
	\frac{1}{\textrm{i} V}
	\frac{ f(\varepsilon_{n \bm k}) - f(\varepsilon_{m \bm k}) }{ \varepsilon_{n \bm k} - \varepsilon_{m \bm k} }
	\nonumber \\
	  &
	\qquad \frac{\braket{n \bm k|j^\mu_{\bm k}|m \bm k} \braket{m \bm k|j^z_{\bm k}|n \bm k}}{\hbar \omega + \varepsilon_{n \bm k} - \varepsilon_{m \bm k} + \textrm{i} \delta}.
	\label{eq:Sigma_MN}
\end{align}
Here, $V$ is the system volume,
$f (\varepsilon)$ is the Fermi-Dirac distribution function,
$\varepsilon_{m \bm k}$ and $\ket{m \bm k}$ are the eigenvalue and eigenstate of $\mathcal H$ for band $m$ with momentum $\bm k$, respectively (we label the bands $m = 1, 2, \cdots, 8$ from the lowest energy to the highest one),
and $j^\mu_{\bm k} = - \partial \mathcal H_{\bm k} / \partial k_\mu$ is the current operator in the $\mu$ direction with momentum $\bm k$, where $\mathcal H_{\bm k}$ is the Fourier component of $\mathcal H$ defined as $\mathcal H = \sum_{\bm{k}} \mathcal H_{\bm k}$.
In the following, we take the elementary charge $e=1$, the Dirac constant $\hbar = 1$,
the temperature $k_\textrm{B} T = 0.1$ ($k_\textrm{B}$ is the Boltzmann constant), and the broadening factor $\delta = 0.02$.

\begin{figure}[b]
	\centering
	\includegraphics[width = 0.9 \columnwidth]{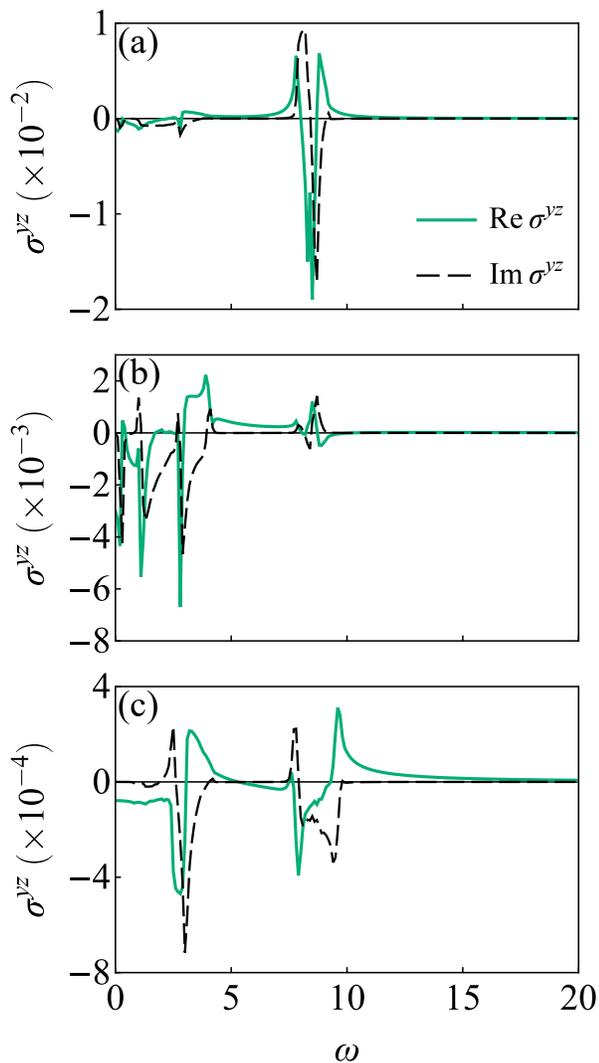}
	\caption{
		\label{fig:TotalSigmaYZ_Square}
		Optical Hall conductivity $\sigma^{yz}(\omega)$ as a function of the energy $\omega$ in the presence of the multipole orders of (a) monopole, (b) quadrupole, and (c) toroidal.
		The green solid (black dashed) line indicates the real (imaginary) part of $\sigma^{yz}(\omega)$.
		The results are computed at the electron filling of $n_\textrm{e} = 0.1$ with $k_\textrm{B}T = 0.1$ and $\delta = 0.02$ in Eqs.~(\ref{eq:KuboWhole}) and (\ref{eq:Sigma_MN}).
		The other model parameters are common to those in Fig.~\ref{fig:Band_Square}.
	}
\end{figure}

In our model, the optical Hall conductivity becomes nonzero for $\sigma^{yz}(\omega)$ in ${\bm B}\parallel [100]$ or $\sigma^{xz}(\omega)$ in ${\bm B}\parallel [010]$ and has the antisymmetric relation $\sigma^{yz}(\omega) = -\sigma^{xz}(\omega)$ from the fourfold rotational symmetry in the $xy$ plane.
We therefore focus on the results of $\sigma^{yz}(\omega)$ in the following.
Figure \ref{fig:TotalSigmaYZ_Square} shows $\sigma^{yz}(\omega)$ as a function of the energy $\omega$ for the (a) monopole, (b) quadrupole, and (c) toroidal orders.
The electron filling $n_\textrm{e} = \frac{1}{2 N}\sum_{i,\sigma} \braket{c^\dagger_{i\sigma}c_{i\sigma}}$ is set to $0.1$
so that the chemical potential lies in the lowest two
bands with the energies $\varepsilon_{1 \bm k}$ and $\varepsilon_{2 \bm k}$ ($N$ is the total number of lattice sites).
We find that the optical Hall conductivity exhibits distinct $\omega$ dependence for different types of the multipole orders.
For the monopole and quadrupole orders, $\sigma^{yz}(\omega)$ shows its primary responses in rather high-energy ($\omega \gtrsim 6$) and low-energy ($0 \leq \omega \lesssim 6$) regions, as shown in Figs.~\ref{fig:TotalSigmaYZ_Square}(a) and \ref{fig:TotalSigmaYZ_Square}(b), respectively.
On the other hand, in the toroidal ordered state, the optical Hall responses in the low- and high-energy regions are comparable to each other, and the overall amplitude is strongly suppressed, as shown in Fig.~\ref{fig:TotalSigmaYZ_Square}(c).

\begin{figure}[b]
	\centering
	\includegraphics[width = 0.9 \columnwidth]{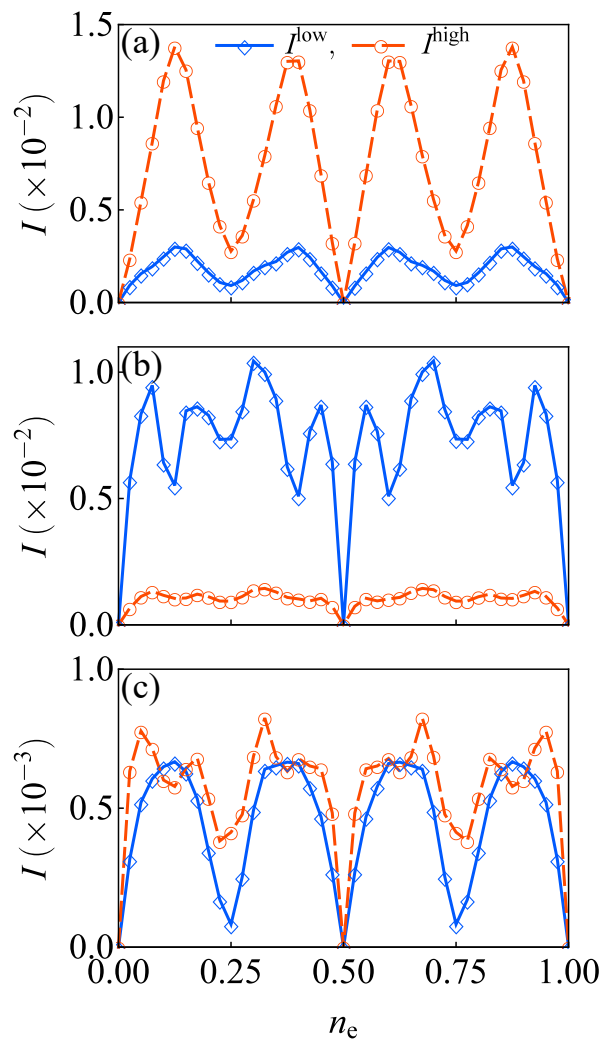}
	\caption{
		\label{fig:IntegralSigmaYZ_Square}
		Integrated intensities of $\left|\text{Re}\, \sigma^{yz}(\omega) \right|$ as functions of the electron filling $n_\textrm{e}$ in the presence of (a) monopole, (b) quadrupole, and (c) toroidal orders [see Eqs.~(\ref{eq:I_low}) and (\ref{eq:I_high})].
		The parameters except for $n_\textrm{e}$ are common to those in Fig.~\ref{fig:TotalSigmaYZ_Square}.
	}
\end{figure}

In order to demonstrate that the distinct responses are generic for any electron filling $n_\textrm{e}$, we compute the integrated intensities $I^\textrm{low}$ and $I^\textrm{high}$ of the absolute values of the real part of $\sigma^{yz}(\omega)$ in the low- and high-energy ranges,
\begin{align}
	\label{eq:I_low}
	I^\textrm{low}
	= &
	\int_{0}^{\omega_1}
	d\omega \,
	\left|\text{Re}\, \sigma^{yz}(\omega) \right| ,
\end{align}
and
\begin{align}
	\label{eq:I_high}
	I^\textrm{high}
	= &
	\int_{\omega_1}^{\omega_2}
	d\omega \,
	\left|\text{Re}\, \sigma^{yz}(\omega) \right|,
\end{align}
respectively, where we take $\omega_1 = 6$ and $\omega_2 = 20$.
Figure~\ref{fig:IntegralSigmaYZ_Square} shows the results as functions of  $n_\textrm{e}$.
They are symmetric with respect to the half filling $n_\textrm{e} =1/2$ because of the particle-hole symmetry between the states of $(k_x, k_y, k_z, \sigma)$ and $(- k_x, - k_y, - k_z+\pi, - \sigma)$.
The optical Hall response vanishes at the half filling as well as empty and full fillings, where the system becomes insulating.
For generic filling, however, $\sigma^{yz}(\omega)$ becomes nonzero.
$I^\textrm{high}$ and $I^\textrm{low}$ are predominant for the monopole and quadrupole orders as shown in Figs.~\ref{fig:IntegralSigmaYZ_Square}(a) and \ref{fig:IntegralSigmaYZ_Square}(b), respectively,
while both responses are comparable to each other and relatively weak for the toroidal order as shown in Fig.~\ref{fig:IntegralSigmaYZ_Square}(c).

\subsection{Decomposition into interband contributions}
\label{sec:InterbandContribution}

\begin{figure}[b]
	\centering
	\includegraphics[width = \columnwidth]{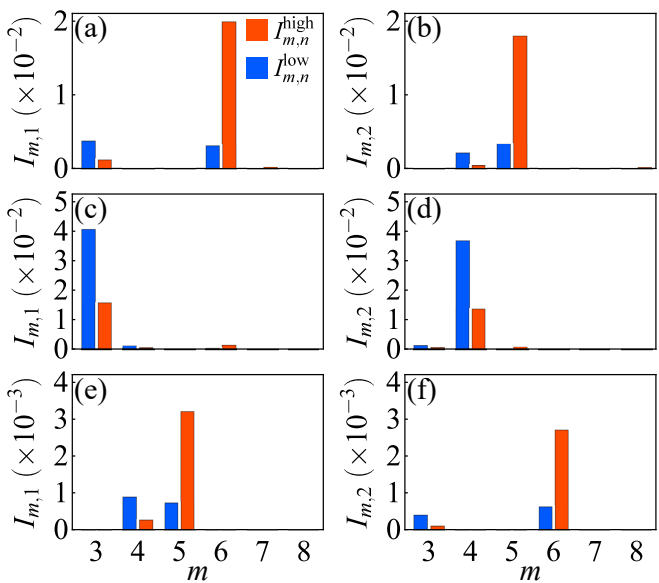}
	\caption{
		\label{fig:BarChart_Square}
		Histogram of the integrated intensities of the interband contributions.
		The blue and red bars represent the low- and high-energy intensities, $I_{m,n}^{\rm low}$ in Eq.~(\ref{eq:I_low_mn}) and $I_{m,n}^{\rm high}$ in Eq.~(\ref{eq:I_high_mn}), respectively, in the presence of (a),(b) monopole, (c),(d) quadrupole, and (e),(f) toroidal orders.
		The results are shown for (a),(c),(e) $n = 1$ and (b),(d),(f) $n = 2$.
		The parameters are common to those in Fig.~\ref{fig:TotalSigmaYZ_Square}.
	}
\end{figure}

In order to clarify which electronic bands play an important role in the optical Hall responses, we decompose the integrated intensities into the interband contributions as
\begin{align}
	I^\textrm{low}_{m,n} = \int_0^{\omega_1} d\omega \,
	\left| \textrm{Re}\, \sigma^{yz}_{m,n}(\omega)\right| ,
  \label{eq:I_low_mn}
\end{align}
and
\begin{align}
	I^\textrm{high}_{m,n} = \int_{\omega_1}^{\omega_2} d\omega \,
	\left| \textrm{Re}\, \sigma^{yz}_{m,n}(\omega)\right| .
  \label{eq:I_high_mn}
\end{align}
We focus on the cases with $n=1,2$ (partially occupied bands) and $m=3,4,\cdots , 8$ (unoccupied bands) at $n_\textrm{e}=0.1$.
The results are plotted in Fig.~\ref{fig:BarChart_Square}.
For the monopole order, the large values of $I^\textrm{high}_{m,n}$ are found for $(m,n) = (6,1)$ and $(5,2)$, as shown in Figs.~\ref{fig:BarChart_Square}(a) and \ref{fig:BarChart_Square}(b), respectively.
Meanwhile, for the quadrupole order, the dominant contributions in $I^\textrm{low}_{m,n}$ appear for $(m,n) = (3,1)$ and $(4,2)$, as shown in Figs.~\ref{fig:BarChart_Square}(c) and \ref{fig:BarChart_Square}(d), respectively.
For the toroidal case shown in Figs.~\ref{fig:BarChart_Square}(e) and \ref{fig:BarChart_Square}(f),
$I^\textrm{low}_{m,n}$ is distributed for $(m,n) = (4,1)$, $(5,1)$, $(3,2)$, and $(6,2)$,
while $I^\textrm{high}_{m,n}$ is concentrated on $(5,1)$ and $(6,2)$.

\begin{figure}[b]
	\centering
	\includegraphics[width = 0.9 \columnwidth]{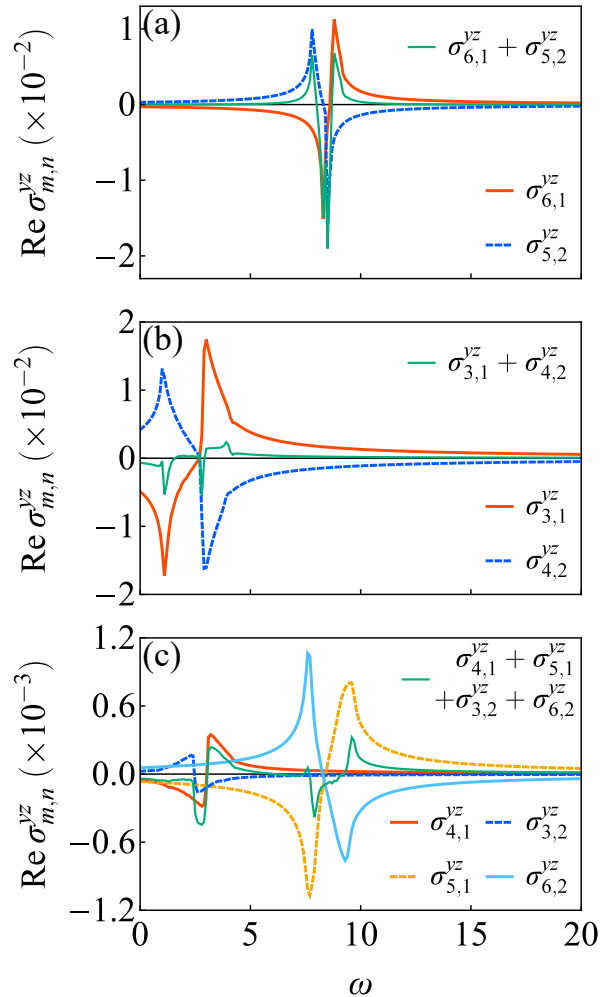}
	\caption{
		\label{fig:Decomposition_Square}
		Dominant interband contributions $\sigma^{yz}_{m,n}(\omega)$ in the presence of (a) monopole, (b) quadrupole, and (c) toroidal orders.
		The parameters are common to those in Fig.~\ref{fig:TotalSigmaYZ_Square}.
	}
\end{figure}

We confirm that the distinct $\omega$ dependences shown in Fig.~\ref{fig:TotalSigmaYZ_Square} are qualitatively explained by these dominant interband contributions.
For the monopole order, as shown in Fig.~\ref{fig:Decomposition_Square}(a), the large response in the high-energy region is well accounted for by the dominant contributions from $(m,n) = (6,1)$ and $(5,2)$.
Similarly, for the quadrupole case, as shown in Fig.~\ref{fig:Decomposition_Square}(b), the large low-energy response is explained by the dominant contributions from $(m,n) = (3,1)$ and $(4,2)$,
in spite of large cancellation between them.
Also in the toroidal ordered state, as shown in Fig.~\ref{fig:Decomposition_Square}(c), the broad and weak response is well reproduced by the dominant interband contributions found in Figs.~\ref{fig:BarChart_Square}(e) and \ref{fig:BarChart_Square}(f).
Thus, the optical Hall responses under different multipole orders originate predominantly from these different interband contributions.

\section{Discussion}
\label{sec:Discussion}

In this section, we discuss the origin of the optical selection rules for the optical Hall conductivity found in the previous section for the different multipole orders.
For this purpose, we approximately estimate $\sigma^{\mu z}_{m,n}(\bm k,\omega)$ in Eq.~(\ref{eq:Sigma_MN}) by examining the matrix elements of the current operators, $\braket{m \bm k | j^{z}_{\bm k} | n \bm k}$ and $\braket{n \bm k|j^y_{\bm k}|m \bm k}$.

Let us begin with the case of the quadrupole order.
First, we consider only the exchange coupling term $\mathcal H_\textrm{MF}$, which has the largest energy scale in our calculations.
We denote the eigenstates of $\mathcal H_\textrm{MF}$ by using the Bloch state with spin $\bm s$ and momentum $\bm k$ at sublattice $l$ as $\ket{\bm s}_{\bm k l}$.
To describe the multipole ordered states with the magnetic moments lying on the $xy$ plane as shown in Figs.~\ref{fig:Structure_Square}(c)-\ref{fig:Structure_Square}(e), we use an arrow for representing the spin direction $\bm s$ in the $xy$ plane,
e.g., $\ket{\searrow}_{\bm k \alpha}$ for the state at sublattice $\alpha$ (lower left) in Fig.~\ref{fig:Structure_Square}(d).
$\mathcal H_\textrm{MF}$ splits the energy levels of the eight Bloch states in a four-site cluster into the fourfold low-energy ones with the eigenenergy of $-M/2$ and the other fourfold high-energy ones with the eigenenergy of $+M/2$.
The eigenstates are given by ($\ket{\searrow}_{\bm k \alpha}$, $\ket{\swarrow}_{\bm k \beta}$, $\ket{\nwarrow}_{\bm k \gamma}$, $\ket{\nearrow}_{\bm k \delta}$) and ($\ket{\nwarrow}_{\bm k \alpha}$, $\ket{\nearrow}_{\bm k \beta}$, $\ket{\searrow}_{\bm k \gamma}$, $\ket{\swarrow}_{\bm k \delta}$), respectively [see Fig.~\ref{fig:Structure_Square}(d)].

Next, we discuss the effect of electron hoppings $\mathcal H_t$ on the four low-energy states.
In the following treatment of $\mathcal{H}_t$ and $\mathcal{H}_{\rm Zeeman}$, we neglect the hybridization between the low-energy and  high-energy states.
By using the basis set of ($\ket{\searrow}_{\bm k \alpha}$, $\ket{\swarrow}_{\bm k \beta}$, $\ket{\nwarrow}_{\bm k \gamma}$, $\ket{\nearrow}_{\bm k \delta}$), $\mathcal H_{t\bm{k}}$, which is defined as $\mathcal H_t = \sum_{\bm{k}} \mathcal H_{t\bm{k}}$, is expressed in the matrix form of
\begin{align}
	\frac{1}{\sqrt{2}}
	\begin{pmatrix}
	0        & \tau_{x}^* & 0            & \tau_{y}^* \\
	\tau_{x} & 0          & - \tau_{y}^* & 0          \\
	0        & - \tau_{y} & 0            & \tau_{x}   \\
	\tau_{y} & 0          & \tau_{x}^*   & 0          \\
	\end{pmatrix}
	+ \tau_z I
	,
	\label{eq:Eff_H_t_+Q}
\end{align}
where
\begin{align}
	\tau_x & = - t_1
	e^{\mathrm{i} k_x \tilde{a}} - t_2
	e^{- \mathrm{i} k_x (a -\tilde{a})} , \\
	\tau_y & = - t_1
	e^{\mathrm{i} k_y \tilde{a}} - t_2
	e^{- \mathrm{i} k_y (a -\tilde{a})} , \\
	\tau_z & = - 2 t_z \cos{(k_z c)} ,
	\label{eq:def:tau}
\end{align}
and $I$ is the $4\times 4$ identity matrix.
The four eigenstates of Eq.~(\ref{eq:Eff_H_t_+Q}) are split into two manifolds, each of which is doubly degenerate.
One has the eigenvalue of $-C_{xy}/\sqrt{2}  + \tau_{z}$ and the eigenstates of
\begin{align}
	\frac{1}{\sqrt{2}}
	\begin{pmatrix}
	\cos{\rho} \,
	e^{-\mathrm{i}\eta} \\
	0                   \\
	\sin{\rho} \,
	e^{\mathrm{i}\phi}  \\
	-1                  \\
	\end{pmatrix}, \
	\frac{1}{\sqrt{2}}
	\begin{pmatrix}
	\sin{\rho} \,
	e^{-\mathrm{i}\phi} \\
	-1                  \\
	-\cos{\rho} \,
	e^{\mathrm{i}\eta}  \\
	0                   \\
	\end{pmatrix},
	\label{eq:+Q_-cxy}
\end{align}
while the other has the eigenvalue of $C_{xy}/\sqrt{2} + \tau_{z}$ and the eigenstates of
\begin{align}
	\frac{1}{\sqrt{2}}
	\begin{pmatrix}
	\cos{\rho} \,
	e^{-\mathrm{i}\eta} \\
	0                   \\
	\sin{\rho} \,
	e^{\mathrm{i}\phi}  \\
	1                   \\
	\end{pmatrix},\
	\frac{1}{\sqrt{2}}
	\begin{pmatrix}
	\sin{\rho} \,
	e^{-\mathrm{i}\phi} \\
	1                   \\
	- \cos{\rho} \,
	e^{\mathrm{i}\eta}  \\
	0                   \\
	\end{pmatrix} .
	\label{eq:+Q_+cxy}
\end{align}
Here, $C_{xy} = \sqrt{|\tau_x|^2 + |\tau_y|^2}>0$, $\tau_x = C_{xy} \sin{\rho}\,e^{\mathrm{i}\phi}$, and $\tau_y = C_{xy} \cos{\rho}\,e^{\mathrm{i} \eta}$ with $\rho \in [0,\frac{\pi}{2}]$, $(\phi, \eta) \in [0,2 \pi)$.
Note that $\mathcal{PT}$ symmetry of $\mathcal{H}_t + \mathcal{H}_{\rm ASOC} + \mathcal{H}_{\rm MF}$ results in the twofold degeneracy in Eqs.~(\ref{eq:+Q_-cxy}) and (\ref{eq:+Q_+cxy}).

The remaining degeneracy is lifted by the Zeeman coupling $\mathcal H_\textrm{Zeeman}$.
In the first-order perturbation, $\mathcal H_\textrm{Zeeman}$ is given by a $2\times 2$ matrix for each Kramers doublet as
\begin{align}
	- \frac{B\cos{\rho}}{2\sqrt{2}}
	\begin{pmatrix}
	\cos{\rho}                     & \sin{\rho} \,
	e^{\mathrm{i}(-\phi + \eta)} \\
	\sin{\rho} \,
	e^{-\mathrm{i}(- \phi + \eta)} & -\cos{\rho}
	\end{pmatrix}.
	\label{eq:Eff_Hz_+Q}
\end{align}
By diagonalizing Eq.~(\ref{eq:Eff_Hz_+Q}), we obtain the eigenstates:
\begin{align}
	{\ket{1 \bm k}}
	=
	\frac{1}{\sqrt{2}}
	\begin{pmatrix}
	\tilde{c}\theta^*_{+}   \\
	- \tilde{s}\theta_{-}   \\
	\tilde{s}\theta_{+}     \\
	- \tilde{c}\theta^*_{-}
	\end{pmatrix},\
	{\ket{2 \bm k}}
	=
	\frac{1}{\sqrt{2}}
	\begin{pmatrix}
	-\tilde{s}\theta^*_{+}  \\
	\tilde{c}\theta_{-}     \\
	\tilde{c}\theta_{+}     \\
	- \tilde{s}\theta^*_{-}
	\end{pmatrix},
	\label{eq:2k0Q}
\end{align}
for Eq.~(\ref{eq:+Q_-cxy}) and
\begin{align}
	{\ket{3 \bm k}}
	=
	\frac{1}{\sqrt{2}}
	\begin{pmatrix}
	\tilde{c}\theta^*_{+}  \\
	\tilde{s}\theta_{-}    \\
	\tilde{s}\theta_{+}    \\
	\tilde{c}\theta^*_{-}
	\end{pmatrix},\
	{\ket{4 \bm k}}
	=
	\frac{1}{\sqrt{2}}
	\begin{pmatrix}
	-\tilde{s}\theta^*_{+} \\
	- \tilde{c}\theta_{-}  \\
	\tilde{c}\theta_{+}    \\
	\tilde{s}\theta^*_{-}
	\end{pmatrix},
	\label{eq:4k0Q}
\end{align}
for Eq.~(\ref{eq:+Q_+cxy}), where $\tilde{c} = \cos(\rho/2)$, $\tilde{s} = \sin(\rho/2)$, and $\theta_{\pm} = \exp[\mathrm{i}(\phi\pm\eta)/2]$.

\begin{figure}[b]
	\centering
	\includegraphics[width = 0.9 \columnwidth]{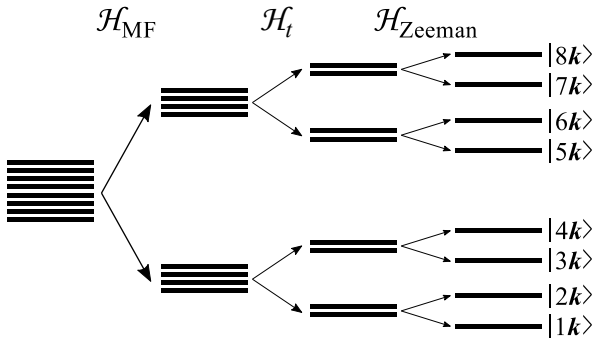}
	\caption{
		\label{fig:BandSplit}
		Schematic picture of the energy levels.
		The eightfold degeneracy in a four-site cluster is lifted successively by $\mathcal H_\textrm{MF}$, $\mathcal H_t$, and $\mathcal H_\textrm{Zeeman}$.
	}
\end{figure}

Similar procedures of the degeneracy lifting by $\mathcal H_\textrm{MF}$, $\mathcal H_t$, and $\mathcal H_\textrm{Zeeman}$ hold for the high-energy four states.
Consequently, the eightfold degenerate states in the four-site cluster are split into $\ket{1 \bm k}, \ket{2 \bm k}, \cdots, \ket{8 \bm k}$, as schematically shown in Fig.~\ref{fig:BandSplit}.
The eight states yield the band structure shown in Fig.~\ref{fig:Band_Square}(b).

Since the chemical potential is set in the lowest two bands ($n=1, 2$) in the present calculations, dominant contributions to $\sigma^{yz}_{m,n}(\bm {k}, \omega)$ come from the matrix elements of $\braket{ m \bm k | j^z_{\bm k} | n\bm k}$ in Eq.~(\ref{eq:Sigma_MN}) for $n = 1$ or $2$, and $m \neq n$.
Among the four terms in Eq.~(\ref{eq:Hamiltonian_All}), only $\mathcal H_t$ and $\mathcal H_\textrm{ASOC}$ contribute to $j^z_{\bm k} = -\partial \mathcal H_{\bm{k}} / \partial \bm{k}$ as
\begin{align}
	j^z_{t\bm{k}}
	= &
	- \frac{\partial {\mathcal H}_{t\bm{k}}}{\partial k_z}
	=
	- 2t_z\sin{(k_z c)}\sum_n{c^\dagger_{n\bm k}c_{n\bm k}} ,
	\label{eq:jzt}
\end{align}
and
\begin{align}
	j^z_{\textrm{ASOC}\bm{k}}
	= &
	- \frac{\partial \mathcal H_{{\rm ASOC}\bm{k}}}{\partial k_z}
	=
	2c\cos(k_zc)\sum_{l}\bm{\Delta}_{l} \cdot \bm {s}_{{\bm k} l} ,
	\label{eq:jzASOC}
\end{align}
respectively, where $\mathcal H_{\rm ASOC} = \sum_{\bm{k}} \mathcal H_{{\rm ASOC} \bm{k}}$ and $\bm{\Delta}_{l}$ is given by $(D_l^y, -D_l^x, 0)$ within the unit cell
[$D_l^\mu$ is the $\mu$ component of $\bm{D}_l$; see Eq.~(\ref{eq:D_l})].
Note that $\bm{\Delta}_l$ is regarded as an effective magnetic field which has a toroidal-like configuration at the four sublattices.
Since $j_{t\bm{k}}^z$ in Eq.~(\ref{eq:jzt}) is diagonal in momentum space, it has only nonzero values for the intraband contributions $\braket{1 \bm k | j^z_{t\bm k} | 1 \bm k}$ and $\braket{2 \bm k | j^z_{t\bm k} | 2 \bm k}$.
On the other hand, $j^z_{\textrm{ASOC}\bm{k}}$ in Eq.~(\ref{eq:jzASOC}) has interband contributions.
The nonzero values of $\braket{m \bm k | j^z_{\textrm{ASOC} \bm{k}} | n \bm k}$ are found only for $n=1$ or $2$ and $1\leq m\leq 4$ within the present approximation because all of the bases $\ket{\searrow}_{\bm k \alpha}$, $\ket{\swarrow}_{\bm k \beta}$, $\ket{\nwarrow}_{\bm k \gamma}$, and $\ket{\nearrow}_{\bm k \delta}$ (the low-energy eigenstates of $\mathcal H_\textrm{MF}$) are the eigenstates of $j^z_{\textrm{ASOC} \bm{k}}$ with the toroidal-like $\bm{\Delta}_l$ which is parallel or antiparallel to the quadrupole order $\bm{M}_l$ at each sublattice.
For the basis set of the four low-energy states, $j^z_{\textrm{ASOC}\bm{k}}$ is written in the matrix form of
\begin{align}
	-2c\cos (k_z c)
	\begin{pmatrix}
	D & 0  & 0 & 0  \\
	0 & -D & 0 & 0  \\
	0 & 0  & D & 0  \\
	0 & 0  & 0 & -D \\
	\end{pmatrix}
	.
	\label{eq:Eff_H_asoc_+Q}
\end{align}
Consequently, in the quadrupole ordered state, we obtain the selection rule for the interband contributions:
\begin{align}
	& \braket{m \bm k | j^z_{\bm k} | n \bm k} \nonumber \\
	& \qquad =
	\begin{cases}
	- 2Dc\cos{(k_z c)} & \text{for} \ (m,n) = (3,1), (4,2) \\
	0                  & \text{otherwise}.
	\end{cases}
	\label{eq:mjzn_selection}
\end{align}

The other matrix element $\braket{n \bm k | j^y_{\bm k} | m \bm k}$ in $\sigma^{yz}_{m,n}(\bm {k}, \omega)$ is also estimated by the same basis set.
Considering Eq.~(\ref{eq:mjzn_selection}), the important contributions are calculated as
\begin{align}
	\braket{ 1 \bm k | j^y_{\bm k} | 3 \bm k }
	=
	- \braket{ 2 \bm k | j^y_{\bm k} | 4 \bm k }
	=
	\frac{\textrm {i}}{2\sqrt{2}}
	\
	\textrm{Im} \left[ \left(
	\frac{\partial \tau_y}{\partial k_y}
	\right)^*
	e^{\textrm{i} \eta} \right].
	\label{eq:njym}
\end{align}
Combining Eqs.~(\ref{eq:mjzn_selection}) and (\ref{eq:njym}), $\sigma^{yz}_{m,n}(\bm k, \omega)$ for the quadrupole ordered state is approximately given as
\begin{align}
& \sigma^{y z}_{m,n}(\bm k,\omega) \nonumber \\
& \ =
	\begin{cases}
	\{ f(\varepsilon_{1 \bm k}) - f(\varepsilon_{3 \bm k}) \} \Xi(\bm k, \omega)   & \text{for} \ (m,n)= (3,1)                  \\
	- \{ f(\varepsilon_{2 \bm k}) - f(\varepsilon_{4 \bm k}) \} \Xi(\bm k, \omega) & \text{for} \ (m,n)= (4,2)                  \\
	0                                                                              & \text{otherwise},
	\end{cases}
\end{align}
where
\begin{align}
	\Xi(\bm k, \omega) =
	\frac{1}{V}
	\frac{Dc\cos{(k_z c)} \ \textrm{Im}[
	(\partial \tau_y/\partial k_y)^*
	e^{\mathrm{i}\eta}]}{{\sqrt{2}}C_{xy} (\hbar \omega - \sqrt{2} C_{xy} +
	{\rm i} \delta)}.
	\label{eq:Xi}
\end{align}
The results explain well the dominant interband contributions found in Sec.~\ref{sec:InterbandContribution}:
the dominant contributions appear only for $(m,n)=(3,1)$ and $(4,2)$ with opposite sign in rather low-energy regions where $\hbar\omega \sim \sqrt{2}C_{xy}$ in the denominator in Eq.~(\ref{eq:Xi}) [see Figs.~\ref{fig:BarChart_Square}(c), \ref{fig:BarChart_Square}(d), and \ref{fig:Decomposition_Square}(b)].
Thus, the optical selection rule for the quadrupole ordered state is rooted in the selection rule of $\bra{m\bm k}j^z_{\bm{k}}\ket{n\bm k}$ in Eq.~(\ref{eq:mjzn_selection}).

Next, we discuss the monopole case.
A difference between the quadrupole and monopole orders lies in the relative angles between the magnetic moments $\bm{M}_l$ and the effective magnetic field $\bm{\Delta}_l$ in Eq.~(\ref{eq:jzASOC}):
while $\bm{M}_l$ is parallel or antiparallel to $\bm{\Delta}_l$ for the quadrupole order, it is perpendicular to $\bm{\Delta}_l$ for the monopole order.
Thus, the eigenstates of $\mathcal{H}_\textrm{MF}$
for the monopole order, $\ket{\swarrow}_{\bm k \alpha}$, $\ket{\searrow}_{\bm k \beta}$, $\ket{\nearrow}_{\bm k \gamma}$, and $\ket{\nwarrow}_{\bm k \delta}$ are spin flipped by $j^z_{{\rm ASOC}{\bm k}}$.
This means that the matrix elements become nonzero for the interband processes with $m$ belonging to the four high-energy levels split by $\mathcal H_{\rm MF}$.
Consequently, the selection rule for this case is given by
\begin{align}
						& \braket{m \bm k | j^z_{\bm k} | n \bm k} \nonumber \\
						& \qquad =
	\begin{cases}
	- 2Dc\cos{(k_z c)} & \text{for} \ (m,n) = (6,1), (5,2)                  \\
	0                  & \text{otherwise}.
	\end{cases}
\end{align}
Following a similar procedure to the quadrupole case above, we end up with
\begin{align}
													& \sigma^{y z}_{m,n}(\bm k,\omega) \nonumber \\
													& \quad =
	\begin{cases}
	f(\varepsilon_{1 \bm k}) \Xi_+(\bm k, \omega)   & \text{for} \ (m,n)= (6,1)                  \\
	- f(\varepsilon_{2 \bm k}) \Xi_-(\bm k, \omega) & \text{for} \ (m,n)= (5,2)                  \\
	0                                               & \text{otherwise},
	\end{cases}
\end{align}
where
\begin{align}
		& \Xi_{\pm}(\bm k, \omega) =
	\frac{1}{V}
	\frac{Dc\cos{(k_z c)} \ \textrm{Im}[
	(\partial \tau_y/\partial k_y)^*
	e^{\mathrm{i}\eta}]}{\sqrt{2}M_\pm (\hbar \omega - M_\pm +
	{\rm i} \delta)}, \\
		& M_\pm = M \pm \frac{1}{\sqrt{2}}B\cos\rho.
\end{align}
Thus, the optical Hall responses in the monopole ordered state appear dominantly in rather high-energy regions corresponding the energy scale of $\mathcal H_{\rm MF}$, namely, $\hbar \omega \sim M_\pm$.
The result explains well again the findings in Figs.~\ref{fig:BarChart_Square}(a), \ref{fig:BarChart_Square}(b), and \ref{fig:Decomposition_Square}(a), as in the case of the quadrupole order.

Finally, in the case of the toroidal order, $\bm M_l$ is in the same direction to $\bm{\Delta}_l$.
This means that $j^z_{{\rm ASOC}{\bm k}}$ is proportional to an identity matrix in the four low-energy eigenstates.
Hence, $j^z_{{\rm ASOC}{\bm k}}$ as well as $ j^z_{t\bm{k}}$ does not lead to any interband excitations, resulting in $\sigma^{yz}_{m,n}(\bm k,\omega>0) = 0$ for all $(m,n)$ within the present approximation.
This explains the small responses found in Figs.~\ref{fig:BarChart_Square}(e), \ref{fig:BarChart_Square}(f), and \ref{fig:Decomposition_Square}(c); they originate in the contributions beyond the present approximation.

Since the optical selection rules discussed here are based on the atomic bases under strong correlation, they are generic to spin-orbit coupled metals under strong influence of the cluster multipole orders, irrespective of the lattice structures and detailed electronic band structures.
To confirm this, we study a honeycomb-lattice variant in Appendix, and obtain optical Hall spectra obeying similar optical selection rules.

\section{Summary}
\label{sec:Summary}
In summary, we have theoretically investigated the optical Hall responses in spin-orbit coupled metals with ferroic orders of cluster-type magnetic multipoles.
Taking a minimal model with monopole, quadrupole, and toroidal orders, we unveiled that the optical Hall conductivity shows distinct frequency dependence for the three types of multipoles.
In the cases of the monopole and quadrupole orders, the predominant response appears in high- and low-energy regions, which correspond to characteristic energy scales of electron correlation and kinetic energy, respectively.
Meanwhile, in the case of the toroidal order, the response is spread over both energy regions with relatively suppressed intensity.
Careful analysis on the interband contributions showed that these distinct optical Hall responses are rooted in the optical selection rules coming from the interplay between the antisymmetric spin-orbit coupling and the underlying cluster multipole ordering.

Our results indicate that the careful investigation of the optical Hall conductivity would be helpful to probe and distinguish the magnetic multipole orders in experiments.
It would also be interesting to extend our study to electric multipoles, which are often more difficult to detect compared to the magnetic ones.
While our model includes the essential ingredients for the spin-orbit coupled metals, further realistic models would be necessary to discuss candidate materials, such as UNi$_{4}$B~\cite{Mentink1994, Hayami2014, Saito2018}, Cd$_{2}$Re$_{2}$O$_{7}$~\cite{Yamaura2002, Yamaura2017, Hiroi2018, Hayami2019}, and PbRe$_{2}$O$_{6}$~\cite{Tajima2020}.
Our work would serve as a starting point for such future studies.

\begin{acknowledgments}
T.S. and Y.S. were supported by the Japan Society for the Promotion of Science through Program for Leading Graduate Schools (MERIT).
Y.S. was also supported by the Japan Society for the Promotion of Science through a research fellowship for young scientists.
This research was supported by Grant-in-Aid for Scientific Research Grants Number JP19H05822 and JST CREST (JP-MJCR18T2).

\end{acknowledgments}
T.S. and Y.U. contributed equally to this work.

\appendix*

\section{Layered honeycomb lattice}
\label{sec:LayeredHoneycombLattice}

\begin{figure}[b]
	\centering
	\includegraphics[width = 0.8 \columnwidth]{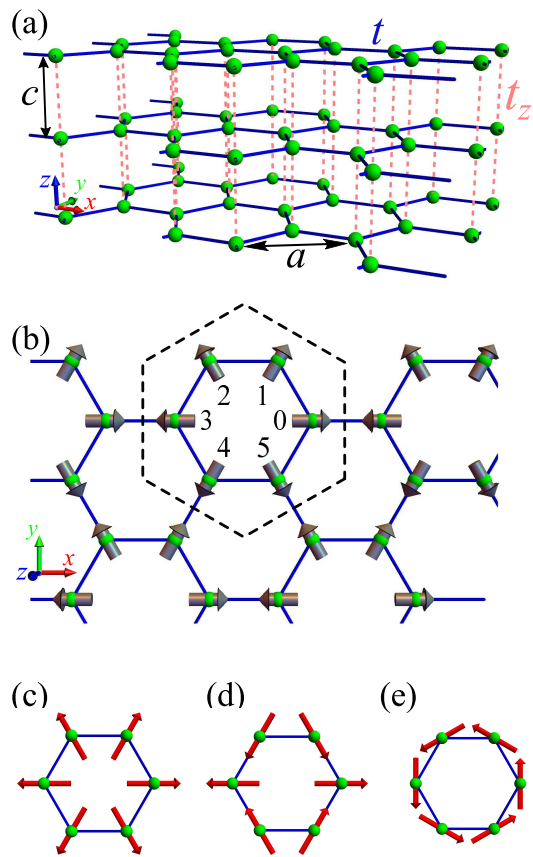}
	\caption{
		\label{fig:Structure_Honeycomb}
		Schematics of (a) a perspective view and (b) a top view of the layered honeycomb lattice.
		In (a), the transfer integrals $t$ and $t_z$ are shown.
		$a$ and $c$ are the lattice constants in and out of the plane, respectively.
		In (b), the dotted hexagon indicates the six-sublattice magnetic unit cell.
	    The gray arrows denote the directions of ${\bm D}_{l}$ at each sublattice; see Eqs.~(\ref{eq:D_l}) and (\ref{eq:Theta^D_l_Honeycomb}).
		(c)-(e) Schematics of cluster-type magnetic multipoles composed of six magnetic dipoles (red arrows): (c) monopole (d) quadrupole-type, and (e) toroidal.
	}
\end{figure}

\begin{figure}[t]
	\centering
	\includegraphics[width = \columnwidth]{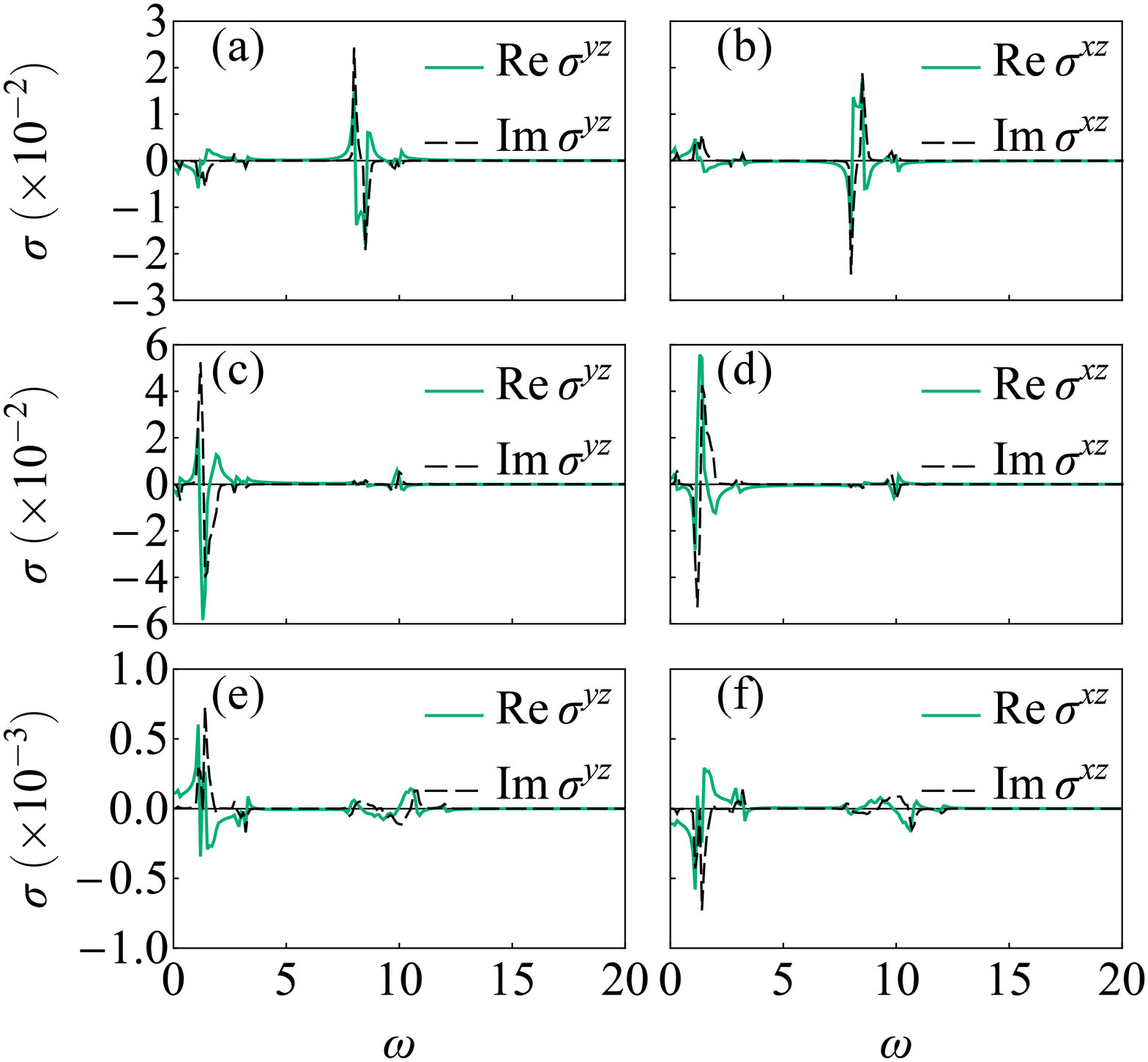}
	\caption{
		\label{fig:TotalSigma_Honeycomb}
		Optical Hall conductivities (a),(c),(e) $\sigma^{yz}(\omega)$ and (b),(d),(f) $\sigma^{xz}(\omega)$ as functions of the energy $\omega$ in the presence of the multipole orders of (a),(b) monopole, (c),(d) quadrupole-type, and (e),(f) toroidal.
    	The green solid (black dashed) line indicates the real (imaginary) part of the optical Hall conductivities.
    	The results are obtained at $a = c = 1$, $t = t_{z} = 1$, $D = 0.5$, $M = 8$, $k_\textrm{B}T = 0.1$, and $\delta = 0.02$.
    	The electron filling is set at $n_\textrm{e} = 0.04$.
    	The magnetic field is applied along the $x$ and $y$ direction for $\sigma^{yz}(\omega)$ and $\sigma^{xz}(\omega)$ as ${\bm B} = (0.5, 0, 0)$ and ${\bm B} = (0, 0.5, 0)$, respectively.
	}
\end{figure}

In order to show the generality of the optical selection rules, we calculate the optical Hall conductivity $\sigma^{\mu z}(\omega)$ in Eq.~(\ref{eq:KuboWhole}) for the layered honeycomb
lattice schematically shown in Fig.~\ref{fig:Structure_Honeycomb}(a).
We adopt a similar Hamiltonian to Eq.~(\ref{eq:Hamiltonian_All}).
In Eq.~(\ref{eq:H_t}), we consider two types of transfer integrals: the intralayer one $t$ and the interlayer one $t_{z}$ [see Fig.~\ref{fig:Structure_Honeycomb}(a)].
All of the other transfer integrals between further-neighbor sites are assumed to be zero.
Note that despite an uniform bond length $a$ and the uniform transfer integral $t$ within each layer, spatial inversion symmetry is broken at each lattice site in the honeycomb case, in contrast to the square case in Fig.~\ref{fig:Structure_Square}.
We choose $\theta^D_l$ in Eq.~(\ref{eq:D_l}) as
\begin{align}
	\theta^D_l
	=
	\frac{\pi}{3} n_{l} ,
	\label{eq:Theta^D_l_Honeycomb}
\end{align}
and $\theta^M_l$ in Eq.~(\ref{eq:M_l}) as
\begin{align}
    & \theta^M_l = \frac{\pi}{3} n_l \quad{\rm for \ monopole},
    \label{eq:theta^M_l_flux_Honeycomb} \\
    & \theta^M_l = - \frac{\pi}{3} n_l  \quad{\mbox{for quadrupole-type}},
    \label{eq:theta^M_l_quadrupole_Honeycomb} \\
	  &	\theta^M_l = \frac{\pi}{3} n_l + \frac{\pi}{2} \quad{\rm for \ toroidal},
    \label{eq:theta^M_l_toroidal_Honeycomb}
\end{align}
where $n_l = $ 0, 1, 2, 3, 4, and 5 identify the six sublattices in the magnetic unit cell surrounded by the dashed hexagon in Fig.~{\ref{fig:Structure_Honeycomb}}(b).
The directions of ${\bm D}_l$ specified by Eq.~(\ref{eq:Theta^D_l_Honeycomb}) are shown by the gray arrows in Fig.~{\ref{fig:Structure_Honeycomb}}(b), and those of ${\bm M}_l$ specified by Eqs.~(\ref{eq:theta^M_l_flux_Honeycomb})-(\ref{eq:theta^M_l_toroidal_Honeycomb}) are displayed by the red arrows in Figs.~\ref{fig:Structure_Honeycomb}(c)-\ref{fig:Structure_Honeycomb}(e), respectively.
The model is an extension of that considered in Ref.~\cite{Hayami2014}.

\begin{figure}[t]
	\centering
	\includegraphics[width = \columnwidth]{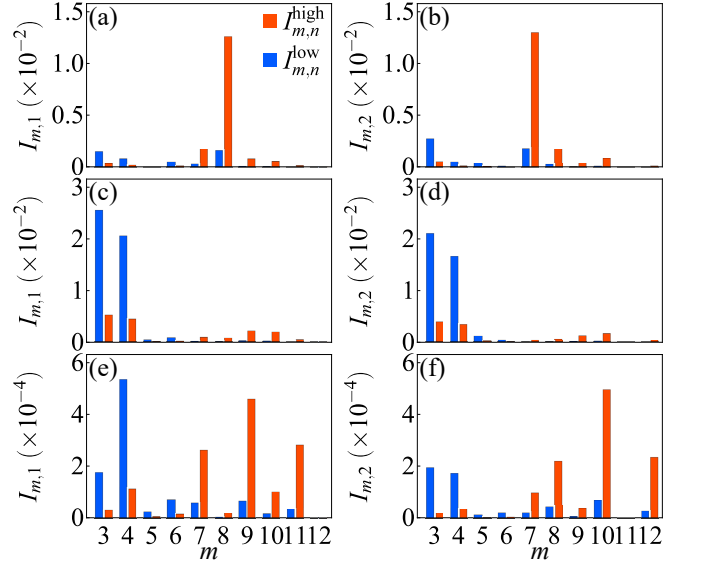}
	\caption{
		\label{fig:BarChart_Honeycomb}
		Histogram of the integrated intensities of the interband contributions $\sigma^{yz}_{m,n} (\omega)$.
		The blue and red bars represent the lower- and higher-energy intensities,	$I_{m,n}^{\rm low}$ in Eq.~(\ref{eq:I_low_mn}) and $I_{m,n}^{\rm high}$ in Eq.~(\ref{eq:I_high_mn}), respectively, in the presence of (a),(b) monopole, (c),(d) quadrupole-type, and (e),(f) toroidal orders.
		The results are shown for (a),(c),(e) $n = 1$ and (b),(d),(f) $n = 2$.
		The parameters are common to those in  Fig.~\ref{fig:TotalSigma_Honeycomb}.
	}
\end{figure}

Figure~\ref{fig:TotalSigma_Honeycomb} shows the optical Hall conductivities $\sigma^{yz} (\omega)$ and $\sigma^{xz} (\omega)$ in a magnetic field applied along the $x$ and $y$ axis, respectively, as functions of the energy $\omega$ for the
(a),(b) monopole, (c),(d) quadrupole-type, and (e),(f) toroidal orders.
The electron filling $n_\textrm{e}$ is set to 0.04 so that the chemical potential lies in the lowest two bands.
We find that $\sigma^{yz} (\omega)$ and $\sigma^{xz} (\omega)$ show distinct $\omega$ dependence for different types of the multipole orders in a similar manner to the square lattice case in Sec.~\ref{sec:OpticalHallConductivity}:
dominant intensities in rather high-energy ($\omega \gtrsim 6$) and low-energy ($0 \leq \omega \lesssim 6$) regions for the monopole and quadrupole orders, respectively, while suppressed responses in both energy regions for the toroidal order.

We also confirm similar optical selection rules by decomposing the integrated intensities into the interband contributions $I_{m,n}^{\rm low}$ and $I_{m,n}^{\rm high}$ with $n = 1, 2$ (partially occupied bands) and $m = 3, 4, \cdots, 12$ (unoccupied bands).
As shown in Fig.~\ref{fig:BarChart_Honeycomb}, the trend is common to those in Fig.~\ref{fig:BarChart_Square}, which supports that similar optical selection rules to those discussed in Sec.~\ref{sec:Discussion} are applicable to this honeycomb case.

\newpage
%\bibliography{main}

%

\end{document}